\documentclass[prd,twocolumn,aps,amsmath,amssymb,nofootinbib,preprintnumbers]
{revtex4}

\usepackage{graphicx}% Include figure files
\usepackage{dcolumn}% Align table columns on decimal point
\usepackage{bm}% bold math
\usepackage{amsmath}
\usepackage{amsfonts}
\def\ls{\mathrel{\lower4pt\vbox{\lineskip=0pt\baselineskip=0pt
           \hbox{$<$}\hbox{$\sim$}}}}
\def\gs{\mathrel{\lower4pt\vbox{\lineskip=0pt\baselineskip=0pt
           \hbox{$>$}\hbox{$\sim$}}}}
\def\drawbox#1#2{\hrule height#2pt

\hbox{\vrule width#2pt height#1pt \kern#1pt
              \vrule width#2pt}
              \hrule height#2pt}

\def\Asym#1#2{\vcenter{\vbox{\drawbox{#1}{#2}
              \kern-#2pt       % line up boxes
              \drawbox{#1}{#2}}}}

\newcommand{\beq}{\begin{equation}}
\newcommand{\eeq}{\end{equation}}

\begin{document}

\title{Gauge invariant MSSM inflaton}

\author{Rouzbeh Allahverdi$^{1,2}$}
\author{Kari Enqvist$^{3}$}
\author{Juan Garcia-Bellido$^{4}$}
\author{Anupam Mazumdar$^{5}$}

\affiliation{$^{1}$~Perimeter Institute for Theoretical Physics, Waterloo, ON,
N2L 2Y5, Canada \\
$^{2}$~Department of Physics and Astronomy, McMaster University, Hamilton,
ON, L8S 4M1, Canada.\\
$^{3}$~Department of Physical Sciences, University of Helsinki,
and Helsinki Institute of Physics,
P.O. Box 64, FIN-00014 University of Helsinki, Finland \\
$^{4}$~Departamento de F\'\i sica Te\'orica \ C-XI, Universidad
Aut\'onoma de Madrid, Cantoblanco, 28049 Madrid, Spain \\
$^{5}$~NORDITA, Blegdamsvej-17, Copenhagen-2100, Denmark}

\date{May 3, 2006}

\begin{abstract}
We argue that all the necessary ingredients for successful inflation
are present in the flat directions of the Minimally Supersymmetric
Standard Model.  We show that out of many gauge invariant
combinations of squarks, sleptons and Higgses, there are two
directions, ${\bf LLe}$, and ${\bf udd}$, which are promising
candidates for the inflaton. The model predicts more than $10^3$
e-foldings with an inflationary scale of $H_{\rm inf}\sim {\cal
O}(1-10)$ GeV, provides a tilted spectrum with an amplitude of
$\delta_H\sim 10^{-5}$ and a negligible tensor perturbation.  The
temperature of the thermalized plasma could be as low as $T_{rh}\sim
{\cal O}(1-10)$~TeV.  Parts of the inflaton potential can be
determined independently of cosmology by future particle physics
experiments.

\end{abstract}
\preprint{NORDITA-2006-13, IFT-UAM/CSIC-06-18, HIP-2006-22/TH, hep-ph/0605035}
\maketitle

%%%%%%%%%%%%%%%%%%%%%%%%%%%%%%%%%%%%%%%%%%%%%%%%%%%%%%%%%%%%%%%%%
\noindent

The one crucial ingredient still missing in the otherwise highly
successful theory of primordial inflation is the connection to
particle physics, and in particular to the Standard Model (SM) or its
extensions. In almost all models of inflation the inflaton is treated
as a SM gauge singlet. The only exception is the large N inflationary
models~\cite{ASKO}, where the gauge invariant quasi-flat directions of
SO(N) are responsible for driving assisted inflation at sub-Planckian
VEVS~\cite{ASSIST}, and sometimes a complete gauge singlet whose
origin and couplings are chosen ad-hoc just to fit the observed
cosmological data~\cite{WMAP} without bothering about the relation to
the observed particle contents of the universe~\cite{LYTH}.

The Minimal Supersymmetric Standard Model (MSSM) is a well motivated
extension of the SM with many cosmological
consequences~\cite{ENQVIST-REV}.  MSSM has nearly $300$ gauge
invariant flat directions made up of squarks, sleptons, and
Higgses~\cite{DRT,GKM}, whose potentials are vanishing in the
supersymmetric limit.  However, they are lifted by a soft
supersymmetry (SUSY) breaking mass term, the trilinear $A$-term and by
non-renormalizable superpotential corrections at scales below the
fundamental scale, which we take to be the Planck scale, $M_{\rm
P}= 2.4\times 10^{18}$~GeV.

Although in the MSSM one usually also relies on a gauge singlet
inflaton, in the present paper we shall show that there are two flat
directions which may serve as a low-scale inflaton; we thus provide
the first example of MSSM inflation occurring at scales well below the
Planck scale and involving a sub-Planckian VEV of the flat direction.
Thus we argue that all the inflationary ingredients are present within
MSSM and do not {\it necessarily} require anything beyond MSSM.  In
particular, we show that the MSSM inflaton is capable of creating the
right amplitude of the scalar perturbations with a tilted spectrum.
Moreover, in the present model certain properties of the
inflaton are in principle testable in future collider experiment
such as Large Hadron Collider (LHC)~\cite{LHC}.

Let us begin by considering a flat direction ${\phi}$ with a
non-renormalizable superpotential term
\beq \label{supot}
W = {\lambda_n \over n}{\Phi^n \over M^{n-3}_{\rm P}}\,,
\eeq
where $\Phi$ is the superfield which contains the flat direction.
Within MSSM all the flat directions are lifted by $n=9$
non-renormalizable operator~\cite{GKM}.  Together with the
corresponding $A$-term and the soft mass term, it gives rise to the
following scalar potential for ${\phi}$:
\beq \label{scpot}
V = {1\over2} m^2_\phi\,\phi^2 + A\cos(n \theta  + \theta_A)
{\lambda_{n}\phi^n \over n\,M^{n-3}_{\rm P}} + \lambda^2_n
{{\phi}^{2(n-1)} \over M^{2(n-3)}_{\rm P}}\,,
\eeq
where $m_\phi$ is the soft SUSY breaking mass for $\phi$. Here $\phi$
and $\theta$ denote the radial and the angular coordinates of the
complex scalar field $\Phi=\phi\,\exp[i\theta]$ respectively, while
$\theta_A$ is the phase of the $A$-term (thus $A$ is a positive
quantity with a dimension of mass). Note that the first and third
terms in Eq.~(\ref{scpot}) are positive definite, while the $A$-term
leads to a negative contribution along the directions where $\cos(n
\theta + \theta_A) < 0$.

The maximum impact from the $A$-term is obtained when $\cos(n \theta +
\theta_A) = -1$ (which occurs for $n$ values of $\theta$).  Along
these directions $V$ has a secondary minimum at \beq \phi = \phi_0
\sim \left(m_\phi M^{n-3}_{\rm P}\right)^{1/n-2} \ll M_{\rm P} \eeq
(the global minimum is at $\phi=0$), provided that
\beq \label{extrem}
A^2 \geq 8 (n-1) m^2_{\phi}\,.
\eeq
At this minimum the curvature of the potential along the radial
direction is $\,+ m^2_{\phi}$ (it is easy to see that the curvature is
positive along the angular direction, too), and the potential reduces
to\footnote{If the $A$ is too large, the secondary minimum will be
deeper than the one in the origin, and hence becomes the true
minimum. However, this is {\it phenomenologically} unacceptable as
such a minimum will break charge and/or color~\cite{DRT}. }: $V \sim
m_\phi^2\phi_0^2 \sim m_{\phi}^2\left(m_{\phi} M^{n-3}_{\rm
P}\right)^{2/(n-2)}$.  Now consider the situation where the flat
direction is trapped in the false minimum $\phi_0$. If its potential
energy, $V$, dominates the total energy density of the Universe, a
period of inflation is obtained. The Hubble expansion rate during
inflation will then be
\beq \label{hubble}
H_{\rm inf} \sim {m_\phi \phi_0 \over M_{\rm P}} \sim
m_\phi \left({m_\phi \over M_{\rm P}}\right)^{1/(n-2)}\,.
\eeq
Note that $H_{\rm inf} \ll m_\phi$. This implies that the potential is
too steep at the false minimum and $\phi$ cannot climb over the
barrier which separates the two minima just by the help of quantum
fluctuations during inflation. The situation is essentially the same
as in the old inflation scenario~\cite{GUTH} with no
graceful exit from inflation.

An interesting but rather trivial observation is that the potential
barrier disappears when the inequality in Eq.~(\ref{extrem}) is
saturated, i.e. when
\beq
\label{cond}
A^2 = 8 (n-1) m^2_\phi\,.
\eeq
Then both the first and second derivatives of $V$ vanish at $\phi_0$,
i.e. $V^{\prime}(\phi_0)=0,~ V^{\prime\prime}(\phi_0)=0$, and the
potential becomes very flat along the {\it real direction}. Around
$\phi_0$ the field is stuck in a plateau with potential energy
\begin{eqnarray}\label{phi0}
&&V(\phi_0) = {(n-2)^2\over2n(n-1)}\,m^2_\phi \phi_0^2\,,\\
&&\phi_0 = \left({m_\phi M^{n-3}_{\rm P}\over
\lambda_n\sqrt{2n-2}}\right)^{1/(n-2)}\,.
\end{eqnarray}
However, although the second derivative of the potential vanishes, the
third does not; instead
\beq \label{3rder}
V^{\prime \prime \prime}({\phi_0}) = 2(n-2)^2
{m^2_\phi \over \phi_0}\,.
\eeq
Around $\phi=\phi_0$ we can thus expand the potential as $V(\phi) =
V(\phi_0) + (1/ 3!)V'''(\phi_0)(\phi-\phi_0)^3$. Hence, in the range
$[\phi_0 - \Delta \phi, \phi_0 + \Delta \phi]$, where $\Delta \phi
\sim H_{\rm inf}^2/V^{\prime\prime\prime}(\phi_0) \sim
\left({\phi}^3_0/M^2_{\rm P}\right) \gg H_{\rm inf}$, the real direction
has a flat potential.

We can now solve the equation of motion for the $\phi$ field in the
slow-roll approximation, $3H\dot\phi
=-(1/2)V'''(\phi_0)(\phi-\phi_0)^2$. Note that the field only feels
the third derivative of the potential.  Thus, if the initial
conditions are such that the flat direction starts in the vicinity of
$\phi_0$ with $\dot\phi\approx 0$, then a sufficiently large number of
e-foldings of the scale factor can be generated. In fact, quantum
fluctuations along the tachyonic direction~\cite{LINDE} will drive the
field towards the minimum. However, quantum diffusion is stronger than
the classical force, $H_{\rm inf}/2\pi > \dot\phi/H_{\rm inf}$, for
\beq \label{drift}
{(\phi_0-\phi) \over \phi_0} \ls \Big({m_\phi\phi_0^2 \over
M_{\rm P}^3}\Big)^{1/2}\,,
\eeq
but from then on,
the evolution is determined by the usual slow roll.

A rough estimate of the number of e-foldings is then given by
\beq \label{efold}
{\cal N}_e(\phi) = \int {H d\phi \over \dot\phi}
\simeq {\phi^3_0 \over 2n(n-1) M^2_{\rm P} (\phi_0-\phi)} ~ ,
\eeq
where we have assumed $V'(\phi) \sim (\phi - \phi_0)^2 V'''(\phi_0)$
(this is justified since $V'(\phi_0) \sim 0, V''(\phi_0)\sim 0$). Note
that the initial displacement from $\phi_0$ cannot be smaller than
$H_{\rm inf}$, due to the uncertainty from quantum fluctuations.

Inflation ends when $\epsilon\sim1$, or
\beq \label{end}
{(\phi_0-\phi) \over \phi_0} \sim
\Big({\phi_0 \over 2n(n-1)M_{\rm P}}\Big)^{1/2}\,.
\eeq
After inflation the coherent oscillations of the flat direction excite
the MS(SM) degrees of freedom and reheat the universe.

Let us now identify the possible MSSM inflaton candidates. Recall
first that the highest order operators which give a non-zero $A$-term
are those with $n=6$. This happens for flat directions represented by
the gauge invariant monomials
\beq
\label{fds}
\phi={\bf L}_i {\bf L}_j {\bf e}_k\,; \quad \phi=
{\bf u}\,{\bf d}_i {\bf d}_j\,.
\eeq
The flatness of the potential require that $i \neq j \neq k$ in the
former and $i \neq j$ in the latter. For $n=6$ and $m_{\phi} \sim 1$
TeV, as in the case of weak scale supersymmetry breaking, we find the
following generic results:\\

\noindent
\begin{itemize}

\item{{\it Sub-Planckian VEVs}:\\ In an effective field theory where
the Planck scale is the cut-off, inflationary potential can be trusted
only below the Planck scale, usually a challenge for the model
building~\footnote{Previous attempts have failed to find models with
sub-Planckian VEVS, see e.g.~\cite{Moroi}. In the first reference the
inflaton is not a gauge invariant combination under MSSM, while in the
second reference, thermalization and supergravity effects are not
properly analyzed.}. In our case the flat direction VEV is
sub-Planckian for the non-renormalizable operator $n=6$, i.e.
$\phi_0\sim 1-3\times10^{14}$~GeV for $m_{\phi}\sim 1-10$~TeV, while
the vacuum energy density ranges $V\sim 10^{34}-10^{38}~({\rm GeV})^4$
(in this paper we assume $\lambda_{n}=1$; generically $\lambda_{n}\leq
1$ but its precise value depends on the nature of high energy
physics).}

\item{{\it Low scale inflation}:\\ Although it is extremely hard to
build an inflationary model at low scales, for the energy density
stored in the MSSM flat direction vacuum, the Hubble expansion rate
comes out as low as $H_{\rm inf}\sim 1-10$~GeV.  It might be possible
to lower the scale of inflation further to the electroweak scale.}

\item{{\it Enough e-foldings}:\\ At low scales, $H_{\rm inf}\sim {\cal
O}(1)$~GeV, the number of e-foldings, ${\cal N}_{\rm COBE}$, required
for the observationally relevant perturbations, is much less than
$60$~\cite{LEACH}. In fact the number depends on when the Universe
becomes radiation dominated (note that full thermalization is not
necessary as it is the relativistic equation of state which matters).

If the inflaton decays immediately after the end of inflation, which
has a scale $V\sim 10^{36}~({\rm GeV})^4$, we obtain ${\cal N}_{\rm
COBE} \sim 47$~\cite{LEACH}. The relevant number of e-foldings could
be greater if the scale of inflation becomes larger. For instance, if
$m_{\phi}\sim 10$~TeV, and $V\sim 10^{38}~({\rm GeV})^4$, we have
${\cal N}_{\rm COBE}\sim 50$. For the MSSM flat direction lifted by
$n=6$ non-renormalizable operators, we obtain the total number of
e-foldings as
\beq \label{totalefold}
{\cal N}_{e} \sim \left({\phi_0^2 \over m_\phi
M_{\rm P}} \right)^{1/2} \sim 10^3 \,,
\eeq
computed from the end of diffusion, see Eq.~(\ref{drift}).
This bout of inflation is sufficiently long to produce
a patch of the Universe with no dangerous relics. Domains
initially closer to $\phi_0$ will enter self-reproduction in
eternal inflation.}

\end{itemize}

Let us now consider adiabatic density perturbations. Despite the low
inflationary scale $H_{\rm inf}\sim {\cal O}(1)$ GeV, the flat
direction can generate adequate density perturbations as required to
explain the COBE normalization. This is due to the extreme flatness of
the potential (recall that $V'=0$), which causes the velocity of the
rolling flat direction to be extremely small.  Thus we find an
amplitude of
\beq
\label{amp}
\delta_{H}\simeq \frac{1}{5\pi}\frac{H^2_{inf}}{\dot\phi}
\sim {m_\phi M_{\rm P}\over \phi_{0}^2}\,{\cal N}_{\rm COBE}^2 \sim 10^{-5}\,,
\eeq
for $m_{\phi}\sim 10^{3}-10^{4}$~GeV, where $\phi_0$ is given by
Eq.~(\ref{phi0}).  In the above expression we have used the slow roll
approximation $\dot\phi\simeq -V'''(\phi_0)(\phi_0- \phi)^2/3H_{\rm
inf}$, and Eq.~(\ref{efold}).  Note the importance of the $n=6$
operators lifting the flat directions ${\bf LLe}$ and ${\bf
udd}$. Higher order operators would have allowed for larger VEVs and a
large $\phi_0$, therefore leaving the amplitude of the perturbations
too low.

The spectral tilt of the power spectrum is not negligible because,
although $\epsilon\sim1/{\cal N}_{\rm COBE}^4\ll 1$, the parameter $\eta =
-2/{\cal N}_{\rm COBE}$ and thus
\begin{eqnarray}
\label{spect}
&&n_s = 1 + 2\eta - 6\epsilon \simeq 1 - {4\over {\cal N}_{\rm COBE}} \sim 0.92\,,\\
&&{d\,n_s\over d\ln k} = - {4\over {\cal N}_{\rm COBE}^2} \sim - 0.002\,,
\end{eqnarray}
which agrees with the current WMAP 3-years' data within
$2\sigma$~\cite{WMAP}, while there are essentially no tensor modes.
Note that the tilt can be enhanced to match the central value of the
WMAP 3-years' data while tuning $\lambda_{n}$ to values lower than
$1$.

Recall that quantum loops result in a logarithmic running of the soft
supersymmetry breaking parameters $m_\phi$ and $A$. One might then
worry about their impact on Eq.~(\ref{cond}) and the success of
inflation. Note however that the only implication is that one must use
the VEV-dependent values of $m_\phi$ and $A$ in Eq.~(\ref{cond}) and
in determining $\phi_0$. The crucial ingredient for a successful
inflation , i.e. having a very flat potential such that
$V^{\prime}\left(\phi_0\right) = V^{\prime \prime}\left(\phi_0\right)
= 0$, will remain unchanged under quantum corrections.

After the end of inflation, the flat direction eventually starts
rolling towards its global minimum. The flat direction decays into
light relativistic MS(SM) particles which obtains kinetic equilibrium
rather quickly~\cite{AVERDI1,AVERDI2} with the largest temperature of
the plasma is given by:
\beq
\label{tmax}
T_{max}\sim \left[V(\phi)\right]^{1/4}\sim (H_{\rm inf}M_{\rm
P})^{1/2}\sim 10^{8}~{\rm GeV}\,.
\eeq
Although the plasma heats up to a large value due to large momenta of
the inflaton decay products, the process of thermalization, which
requires chemical equilibrium, can be a slow process~\cite{AVERDI2}.
Just to illustrate, we note that within MSSM there are other flat directions
orthogonal to Eq.~(\ref{fds}) (for an algorithm finding such
multi-directions, see~\cite{JOKINEN}). These can develop large
VEVs and induce large masses to the MSSM quanta, i.e. squarks and
sleptons and gauge bosons and gauginos. As a consequence, there
will be a kinematical blocking for the inflaton to
decay~\cite{ROUZ-REHEAT} which can delay thermalization as
discussed in Refs.~\cite{AVERDI1,AVERDI2}.

%For instance, ${\bf LLe}$ breaks the
%$SU(2)_{L}\times U(1)_y$ during and after inflation, giving masses to
%the corresponding gauge bosons and gauginos, while ${\bf udd}$ gives
%masses to the color degrees of freedom.

%If and only if the initial temperature is very high (higher than the
%masses of squarks, sleptons from the vacuum), then the MSSM quanta
%obtain large thermal mass corrections. This will leave only few light
%degrees of freedom which would thermalize instantly. In that case the
%final reheat temperature will be much lower than that of
%Eq.~(\ref{tmax}).

The details of thermalization would require involved calculations
which are beyond the scope of the present paper. However, perhaps the
best guess is to assume that the flat direction mass gives the lower
limit on a temperature, where all the MSSM degrees of freedom are in
thermal equilibrium~\cite{AVERDI2}:
\beq
T_{rh}\sim m_{\phi}\sim {\cal O}(1-10)~{\rm TeV}\,.
\eeq
Note that this temperature is sufficiently high for electroweak
baryogenesis~\cite{BARYO-REV} and for both thermal and non-thermal
cold dark matter production~\cite{CDM-REV}. Affleck-Dine
baryogenesis~\cite{Affleck-Dine} via other MSSM flat directions is
also an option. For an example, the fragmentation of ${\bf udd}$ can
generate $Q$-balls~\cite{qballs}, which could be responsible for
baryogenesis; and the remnant $Q$-balls could act as cold dark matter
~\cite{ENQVIST-REV}.

Let us also briefly mention an alternative possibility for creating
the density perturbations through a curvaton~\cite{SLOTH}.  MSSM flat
directions can provide us with curvaton candidates~\cite{KASUYA} when
a flat direction orthogonal to the inflaton, $\varphi$, achieves a
large VEV during inflation (provided their potential follows
Eq.~(\ref{scpot}) with $\phi\rightarrow \varphi$ while satisfying
Eq.~(\ref{cond})). If the VEV is such that the amplitude of the
perturbations is $\delta={H_{\inf}}/{\varphi_{\ast}}\sim 10^{-5}$, and
the fluctuations are Gaussian, then such an orthogonal direction can
act as a curvaton.  A promising curvaton candidate is ${\bf H}_u {\bf
Q}_2 {\bf u}_2$ (indices refer to the generation), if ${\bf LLe}$ is
the inflaton.

Let us finally turn to the issue of initial conditions. The Universe
could begin very hot, with all the relevant degrees of freedom
excited.  As it cools, in a small patch it can achieve an initial
state where the flat direction potential is given by Eq.~(\ref{cond})
and inflation can begin. This is the simplest possibility, but there
are also others, e.g. a running scale/coupling could modify the VEV of
$A$ so that the universe is initially stuck in the inflationary
minimum at $\phi_0$, which is later lifted to become a turning point;
or there could be multiple phases of inflation, as has been often
argued, see e.g. Ref.~\cite{BURGESS}, with a last phase driven by the
MSSM inflaton.

To summarize, for the first time a gauge invariant inflaton candidate
have been proposed within MSSM without any inclusion of gauge
singlets. The inflaton candidates are the flat directions ${\bf udd}$
and ${\bf LLe}$.  Both are lifted by non-renormalizable operators at
$n=6$ level. Lower dimension operators do not provide the $A$-term
required for inflation, while higher order operators fail to generate
the observed CMB fluctuations. The model predicts low scale inflation,
non-detectable gravity waves, and a slight departure from scale
invariance. As we have discussed, the conservative estimate for the
spectral index is $n_s\sim 0.92$ but a small enhancement giving a
match to the WMAP 3-year data is possible with a slight tuning of the
flat direction coupling strength. Further note that supergravity
effects are negligible for $H_{\rm inf}\ll m_{\phi}$ and therefore do not
spoil the predictions. More importantly, there will be no moduli problem in
our model. Since $H_{\rm inf} \ll m_\phi$, all moduli will settle at their
true minimum during inflation even if their initial displacement is
${\cal O}(M_{\rm P})$.

The salient feature of the present model is that in principle the
properties of the inflaton are now related to the dynamics of
baryogenesis and the properties of dark matter. Moreover, some
properties of the inflaton potential, such as the soft mass term and
the A-term, could be determined independently of cosmology by particle
physics experiments, possibly already at LHC.

%%%%%%%%%%%%%%%%%%%%%%%%%%%%%%%%%%%%%%%%%%%%%%%%%%%%%%%%%%%%%%%%%%%%%%%%%%%%%%%
{\it Acknowledgments-} We wish to thank Andrew Liddle and Jaume
Garriga for helpful comments. The work of R.A. is supported by the
National Sciences and Engineering Research Council of Canada
(NSERC). J.G.B. is supported in part by a CICYT project
FPA2003-04597. K.E. is supported in part by the Academy of Finland
grant no. 205800.
%%%%%%%%%%%%%%%%%%%%%%%%%%%%%%%%%%%%%%%%%%%%%%%%%%%%%%%%%%%%%%%%%%%%%%%%%%%%%%%

\end{document}